\documentclass[prb,twocolumn,showpacs,floatfix,preprintnumbers,amsmath,amssymb,superscriptaddress]{revtex4}

\usepackage{graphicx}
\usepackage{dcolumn}
\usepackage{bm}
\usepackage{color}

\newcommand{\lao}{LaFeAsO}
\newcommand{\laf}{LaFeAsO$_{1-x}$F$_x$}

\newcommand{\tn}{$T_{\rm{N}}$}
\newcommand{\ts}{$T_{\rm{S}}$}

\newcommand{\etal}{{\it et al.}}

\usepackage{color}

\usepackage[latin1]{inputenc}

\begin{document}

\preprint{APS/123-QED}

\title{Thermal expansion of LaFeAsO: Evidence for high-temperature fluctuations}

\author{L.~Wang}\affiliation{Institute for Solid State Research, IFW Dresden, D-01171 Dresden, Germany}
\author{U.~K{\"o}hler}\email{u.koehler@ifw-dresden.de}\affiliation{Institute for Solid State Research, IFW Dresden, D-01171 Dresden, Germany}
\author{N.~Leps}\affiliation{Institute for Solid State Research, IFW Dresden, D-01171 Dresden, Germany}
\author{A.~Kondrat}\affiliation{Institute for Solid State Research, IFW Dresden, D-01171 Dresden, Germany}
\author{M.~Nale}\affiliation{Van der Waals-Zeeman
Institute, University of Amsterdam, 1018 XE Amsterdam, The
Netherlands}
\author{A.~Gasparini}\affiliation{Van der Waals-Zeeman Institute, University of Amsterdam, 1018 XE Amsterdam, The
Netherlands}
\author{A.~de~Visser}\affiliation{Van der Waals-Zeeman
Institute, University of Amsterdam, 1018 XE Amsterdam, The
Netherlands}
\author{G.~Behr}\affiliation{Institute for Solid State Research, IFW Dresden, D-01171 Dresden, Germany}
\author{C.~Hess}\affiliation{Institute for Solid State Research, IFW Dresden, D-01171 Dresden, Germany}
\author{R.~Klingeler}\affiliation{Institute for Solid State Research, IFW Dresden, D-01171 Dresden, Germany}
\author{B.~B\"uchner}\affiliation{Institute for Solid State Research, IFW Dresden, D-01171 Dresden, Germany}

\date{\today}

\begin{abstract}
We present measurements of the thermal expansion coefficient
$\alpha$ of polycrystalline LaFeAsO$_{1-x}$F$_x$ ($x \leq 0.1$).
The magnetic and structural transitions of the samples with $x
\leq 0.04$ give rise to large anomalies in $\alpha(T)$, while the
onset of superconductivity in the crystals with $x \geq 0.05$ is
not resolved. Above the structural transition, the thermal
expansion coefficient of LaFeAsO is significantly enhanced. This
is attributed to fluctuations, which also affect the electrical
transport properties of the compound. The complete absence of
these fluctuations in the superconducting samples even for $x =
0.05$ is taken as evidence for an abrupt first-order type of
suppression of the structural and magnetic transitions upon F
doping.

\end{abstract}

\maketitle

\section{Introduction}

The family of layered $\rm FeAs$-materials has been extensively
studied since the discovery of superconductivity with transition
temperatures $T_\mathrm{C}$ up to 28~K in \laf
.~\cite{Kamihara2008} In the meantime, $T_\mathrm{C}$ has been
increased to above
50~K~\cite{Chen2008a,Cheng2008,Liu2008a,Ren2008b,Ren2008c} by
replacing La with other rare earths. Superconductivity has also
been found in several related materials, such as
Ba$_{1-x}$K$_x$Fe$_2$As$_2$,~\cite{Rotter2008b}
LiFeAs,~\cite{Pitcher2008,Tapp2008} or FeSe.~\cite{Hsu2008}
Interestingly, the evolution of superconductivity in \laf~seems to
be related to the suppression of a magnetically ordered
orthorhombic phase, which has been found in the undoped parent
compound.~\cite{Luetkens2008,Luetkens2009} In \lao , long range
magnetic order probably of a spin density wave (SDW) type evolves
below 137~K,~\cite{Cruz2008,Klauss2008} while an orthorhombic
distortion of the tetragonal high temperature phase has been
observed at 156~K. The SDW ground state has also been established
for isostructural Rare Earth (R) based $\rm
RO_{1-x}F_xFeAs$,~\cite{Drew2008a} and even in the parent
materials of other iron-pnictide superconductors such as $\rm
Ba_{1-x}K_xFe_2As_2$ with a different crystal structure but
similar $\rm Fe_2As_2$
layers.~\cite{Rotter2008a,Rotter2008b,Rotter2008} However, while
the structural and magnetic phase transitions are separated by
about 20~K in the RFeAsO systems, they coincide for compounds of
AFe$_2$As$_2$ type.

In this paper, we investigate polycrystalline LaFeAsO$_{1-x}$F$_x$
($x \leq 0.1$) by means of thermal expansion measurements which
sensitively probe the volume changes of the material. Large
anomalies of the coefficient of linear thermal expansion $\alpha$
are found at the structural and magnetic transitions of the
samples with $x \leq 0.04$. For comparison we studied
superconducting LaFeAsO$_{1-x}$F$_x$ with $x = 0.05,0.1$, which
exhibits neither the structural nor the magnetic phase transition.
Interestingly, strong differences between the $\alpha (T)$ curves
for the magnetic and the superconducting samples extend to
temperatures well above the structural transition. We analyze
these findings in view of our specific heat, X-ray diffraction and
resistivity data. Our data provide clear evidence for strong
fluctuations in LaFeAsO$_{1-x}$F$_x$ ($x \leq 0.04$) over a large
$T$ range above the structural transition temperature
$T_\mathrm{S}$. By contrast, no indication of fluctuations is
found in the superconducting samples. We discuss the implications
for the phase diagram of LaFeAsO$_{1-x}$F$_{x}$, particularly at
the crossover from a magnetic to a superconducting ground state.


\section{Experimental Methods}

The preparation and characterization of our polycrystalline
samples has been described in detail in
Ref.~\onlinecite{Kondrat2008-0811.4436}. For the thermal expansion
measurement a three-terminal capacitance dilatometer was utilized,
which allows an accurate study of crystal length changes. To be
specific, we measured the macroscopic length $L(T)$ of the samples
and calculated the coefficient of linear thermal expansion $\alpha
= 1/L \cdot dL/dT$, which is the first temperature derivative of
$L(T)$. The specific heat was studied in a Quantum Design PPMS by
means of a relaxation technique. Electrical resistivity
measurements were performed using a standard four-probe technique.


\section{Results}

\begin{figure}[tb]
\begin{center}
\includegraphics[width=0.48\textwidth]{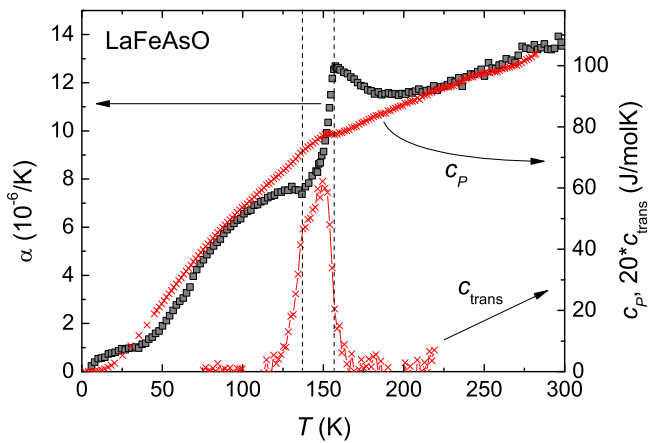}
\end{center}
\caption{Temperature dependence of the coefficient of linear
thermal expansion, $\alpha (T)$, of LaFeAsO in comparison to the
specific heat, $c_p(T)$, of this compound. Two subsequent phase
transitions are found in both quantities at similar temperatures,
as indicated by the vertical lines. In addition, the specific heat
contribution arising from the phase transitions,
$c_{\mathrm{trans}}$, is shown on a larger y scale. \label{texp}}
\end{figure}

\begin{figure}[tb]
\begin{center}
\includegraphics[width=0.43\textwidth]{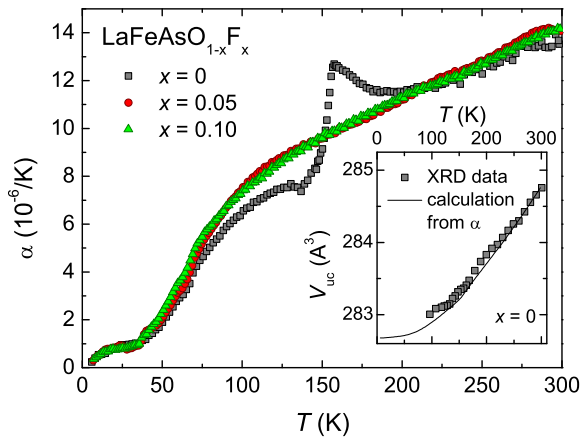}
\end{center}
\caption{Coefficient of linear thermal expansion, $\alpha$,
vs.~temperature, $T$, of LaFeAsO$_{1-x}$F$_x$ for different
fluorine content $x$. For clarity, not all investigated samples
are shown. The inset compares the temperature dependence of the
unit-cell volume, $V_{\mathrm{uc}}(T)$, for $x=0$ obtained from
XRD \cite{Kondrat2008-0811.4436} to the one calculated from
$\alpha$ using the XRD value at 300~K as initial value. The
orthorhombic unit cell is used in the whole $T$ range.
\label{vol}}
\end{figure}

Figure~\ref{texp} shows the linear thermal expansion coefficient
$\alpha$ of LaFeAsO between 5~K and 300~K. In the whole
investigated temperature range, $\alpha$ is found to be positive.
This is in agreement with X-ray diffraction (XRD) data, which
revealed a monotonically increasing lattice volume $V_\mathrm{uc}$
upon heating.~\cite{Kondrat2008-0811.4436} For our polycrystalline
samples the volume expansion coefficient $\beta$ is given as
$\beta = 3 \alpha$. Taking the volume at 300~K as an initial value
we calculated the temperature dependence $V_\mathrm{uc}(T)$, which
agrees with the unit-cell volume determined from XRD data
published in \onlinecite{Kondrat2008-0811.4436} (cf. inset of
Fig.~\ref{vol}a). Below 30 K a small plateau is seen in our
$\alpha (T)$ data, which is also present for the F-doped samples
discussed below. The origin of this structure is unclear. However,
it does not affect the discussion of our data, which focusses on
the temperature region above 100~K.

In Fig.~\ref{texp}, $\alpha (T)$ of LaOFeAs is shown in comparison
to specific heat data $c_p(T)$. Below 120~K and above 200~K,
$\alpha$ roughly tracks the temperature dependence of $c_p$,
indicating an only weakly $T$-dependent Gr\"{u}neisen ratio
$\alpha / c_p$. This observation implies a single energy scale in
this temperature range and agrees to the assumption that mainly
phonon degrees of freedom are relevant. Around 150\,K both
quantities exhibit anomalous contributions, which do not obey a
Gr{\"u}neisen scaling. The thermal expansion coefficient exhibits
two huge anomalies with opposite sign, while the specific heat
evolves rather smoothly with a small additional contribution, as
will be discussed below. The anomalies in $\alpha (T)$ can be
attributed to the structural and SDW transitions of the compound.
The transition temperatures determined from the positions of the
extrema are 137~K and 157~K, respectively. These values are close
to those found in many other properties on the same samples, such
as XRD, resistivity, magnetic
susceptibility,~\cite{Kondrat2008-0811.4436} $\mu$SR, and
M{\"o}ssbauer spectroscopy.~\cite{Klauss2008,Luetkens2009} Both
transitions are also visible in the specific heat data, which
signals anomalous entropy changes in this temperature regime. The
anomalous specific heat $c_\mathrm{trans}$ can be determined by
subtracting the phononic and electronic background estimated from
a polynomial approximation of the data well above ($T
>170$~K) and below ($T <120$~K) the transitions.~\cite{klingeler02} The result of such a procedure
is shown in Fig.~\ref{texp} on an enhanced scale, where the two anomalies at \tn\ and \ts ,
respectively, are clearly visible.

The SDW formation at $T_\mathrm{N}$ = 137 K generates a negative
anomaly in the thermal expansion coefficient. For a second order
phase transition, the slope $\mathrm{d}T_\mathrm{N}/\mathrm{d}p$
can be determined from the jump height at $T_\mathrm{N}$ in the
specific heat $\Delta c_{p}$ and the volume thermal expansion
coefficient $\Delta \beta = 3 \Delta \alpha$ using the Ehrenfest
relation
\begin{equation}\label{eq2}
    \frac{\partial T_\mathrm{N}}{\partial p} = TV_\mathrm{mol} \frac{\Delta \beta}{\Delta c_{p}}
\end{equation}
with the molar volume $V_\mathrm{mol}$. However, for the given
compound a determination of $\Delta c_{p}$ and $\Delta \alpha$ is
not possible with satisfactory accuracy, due to the proximity of
the structural and SDW transitions. Nevertheless, according to
Equation~\ref{eq2} the negative anomaly in $\alpha (T)$ at
$T_\mathrm{N}$ qualitatively clearly implies a negative
hydrostatic pressure dependence of \tn . This finding is in line
with resistivity investigations on LaFeAsO showing a lowering of
$T_\mathrm{N}$ under pressure with an initial slope of
$\mathrm{d}T_\mathrm{N}/\mathrm{d}p|_{p=0} \approx -
1.5$~K/kbar.~\cite{F4-08-7}

Contrary to the magnetic ordering, the structural transition at
$T_\mathrm{S}$ = 157~K gives rise to a positive anomaly in
$\alpha(T)$. This anomaly is very broad, extending to temperatures
far above $T_\mathrm{S}$. Extrinsic effects, in particular those
originating from grain boundaries, cannot cause this broadening.
The grain size of our polycrystalline material has been determined
from electron microscopy to be some tens of
micrometer.\cite{Kondrat2008-0811.4436} This renders a major
contribution from grain boundaries to the thermal expansion
coefficient as observed occasionally in nanocrystalline
material\cite{O-92-1} unlikely. Moreover, while the grain size is
rather unaffected by F-doping the width of the anomaly at
\ts~changes systematically upon substitution of O by F, as shown
below, which also confirms the intrinsic nature of the broadening.
By contrast, the corresponding anomaly in $c_p$ is sharp, which
excludes also the possibility of a smeared transition, e.g.~due to
sample inhomogeneities, as origin of the broadening. A sharp
anomaly is likewise visible in our previously published $\partial
(\chi T)/\partial T$ data.~\cite{Klauss2008} Specific heat,
magnetization, and XRD data indicate a second order phase
transition at \ts . Therefore, one expects a jump in the thermal
expansion coefficient at the phase transition. From our data it is
however not possible to determine the anomalous volume changes at
the transition itself so that an analysis even of the sign of the
pressure dependence $\mathrm{d}T_\mathrm{S}/\mathrm{d}p$ is hardly
possible.

\begin{figure}[tb]
\begin{center}
\includegraphics[width=0.43\textwidth]{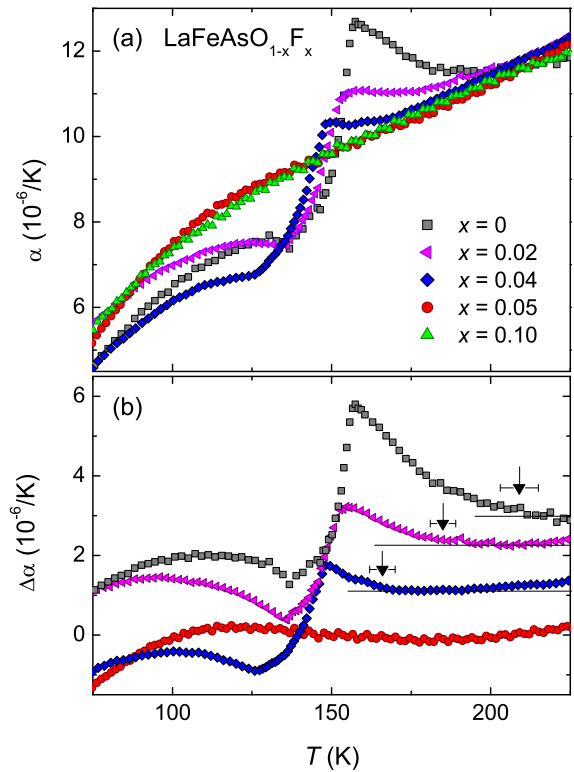}
\end{center}
\caption{(a) Enhanced view of the linear thermal expansion
coefficient vs.~temperature, $\alpha(T)$, of LaFeAsO$_{1-x}$F$_x$
between 80~K and 220~K. (b) Determination of the temperature
$T_\mathrm{fl}$, below which the thermal expansion data point at
the presence of fluctuations. For this purpose, the data for $x =
0.1$ have been subtracted and the curves shifted by multiples of
$10^{-6}/$K. Arrows with error bars mark $T_\mathrm{fl}$.
\label{Th-exp5}}
\end{figure}

In order to understand the behavior of $\alpha(T)$ of LaFeAsO
above $T_\mathrm{S}$ we studied the linear thermal expansion
coefficient of LaFeAsO$_{1-x}$F$_x$ with nominal \mbox{$x = 0.02,
0.04, 0.05, 0.1$}. Upon fluorine substitution for $x\leq 0.04$ the
structural and magnetic phase transitions are only weakly
affected, i.e.~they are slightly shifted to lower temperatures. At
higher F content $x \geq 0.05$ both transitions are completely
suppressed, and a superconducting ground state is
found.~\cite{Luetkens2009} Thus, F doping strongly affects the
electronic properties of the series, especially at low
temperatures, as well as the structural properties below \ts . By
contrast, the properties of the atomic lattice above
$T_\mathrm{S}$, such as the phonon spectrum, are expected to be
relatively insensitive to the fluorine content. Figure~\ref{vol}
compares the coefficient of linear thermal expansion of
LaFeAsO$_{1-x}$F$_x$ for different F content $x = 0, 0.05, 0.1$.
An enhanced view of the $T$ region between 80~K and 220~K for all
investigated F concentrations is shown in Fig.~\ref{Th-exp5}a.
Upon F doping with $x \leq 0.04$, the anomalies arising from the
magnetic and structural transitions are shifted to lower
temperatures, as expected from the lowering of \ts~and \tn . For
$x\geq 0.05$ these transitions are suppressed. At high ($T
> 210$~K) and low ($T \lesssim 70$~K) temperatures, $\alpha$ is almost
independent of the composition. In particular, the superconducting
transitions at $T_{\rm{C}} = 20.6$~K ($x = 0.05$) and $T_{\rm{C}}
= 26.8$~K ($x = 0.1$), which are clearly visible in the
resistivity, are not seen in $\alpha(T)$. Although our data in
this temperature range are influenced by the small plateau of
unknown origin, the existence of large anomalies at $T_{\rm{C}}$
appears unlikely given the close agreement of the curves for
different F content. Regarding the $T$ range above the structural
transition of LaFeAsO$_{1-x}$F$_x$, Fig.\ref{Th-exp5}a clearly
reveals a significantly enhanced $\alpha$ for the magnetic samples
compared to the superconducting ones. This difference cannot be
explained by a simple change of the phonon spectrum upon F doping
due to the larger atomic mass of fluorine compared to oxygen,
since the samples with x = 0.05 and 0.1 exhibit almost identical
$\alpha(T)$ curves. Instead, the enhanced $\alpha$ for $x \leq
0.04$ suggests the presence of strong fluctuations preceding the
structural transitions at $T_\mathrm{S}$.

\begin{figure}[tb]
\begin{center}
\includegraphics[width=0.43\textwidth]{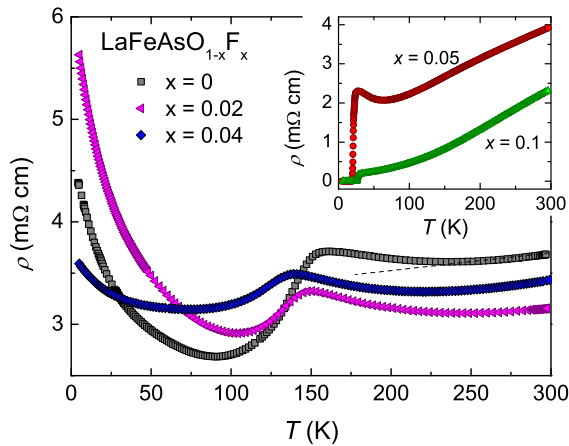}
\end{center}
\caption{Temperature dependence of the resistivity, $\rho$, of
LaFeAsO$_{1-x}$F$_x$. The main plot shows the data for the
magnetic samples with $x\leq0$, while the superconducting samples
($x\geq 0.05$) are shown in the inset. A dashed line to the eye
highlights the regime with positive slope
$\mathrm{d}\rho/\mathrm{d}T$ for $x = 0$. \label{rho}}
\end{figure}

Indications for fluctuations are also found in the electrical
transport properties of LaFeAsO$_{1-x}$F$_x$. Figure~\ref{rho}
shows the electrical resistivities $\rho(T)$ for $x \leq 0.1$
taken from Ref.~\onlinecite{Hess2008}. At room temperature all
samples exhibit metal-like resistivities with a positive slope
$\mathrm{d}\rho/\mathrm{d}T$. Upon cooling, the resistivities of
the superconducting samples continue to decrease, except for an
upturn just above the superconducting transition for $x=0.05$, the
origin of which is still not clear. Transitions to a
superconducting state are found at $T_{\rm{C}} = 20.6$~K ($x =
0.05$) and $T_{\rm{C}} = 26.8$~K ($x = 0.1$). By contrast, the
resistivities of LaFeAsO$_{1-x}$F$_x$ ($x\leq 0.04$) increase
below approximately 230~K and reach maxima at $T_\mathrm{S}$. This
negative temperature dependence indicates enhanced scattering of
charge carriers from fluctuations above $T_\mathrm{S}$.
Consistently, below the transition, $\rho$ drops as the
fluctuations die away.

In Fig.~\ref{phase} we plot the characteristic temperatures
obtained from our thermal expansion measurements in the phase
diagram of LaFeAsO$_{1-x}$F$_x$ established from magnetic
susceptibility, $\mu$SR and resistivity
experiments.\cite{Klingeler2008,Luetkens2009,Hess2008} The values
for \ts~and \tn~taken as the positions of the extrema in
$\alpha(T)$ fit well into this phase diagram. In addition we plot
the temperature $T_\mathrm{fl}$, below which indications for
fluctuations are found in the thermal expansion coefficent. For
this purpose the data for $x = 0.1$ have been subtracted from
$\alpha(T)$ for $x \leq 0.05$. The results are shown in
Fig.~\ref{Th-exp5}b as $\Delta \alpha = \alpha (x) - \alpha (0.1)$
vs.~$T$, whereas the curves have been shifted by multiples of
$10^{-6}/$K. For $x \leq 0.04$, $T_\mathrm{fl}$ is determined as
the temperature, below which $\Delta \alpha$ exhibits a negative
slope. With increasing fluorine content, $T_\mathrm{fl}$ is
lowered. Extrapolating $T_\mathrm{fl}(x)$ and \ts$(x)$ linearly to
higher $x$ would suggest a vanishing of the fluctuation regime
around $x = 0.06$. Nevertheless, already for the sample with $x =
0.05$ no indication of fluctuations is found in $\alpha(T)$.

\section{Discussion}

\begin{figure}[tb]
\begin{center}
\includegraphics[width=0.43\textwidth]{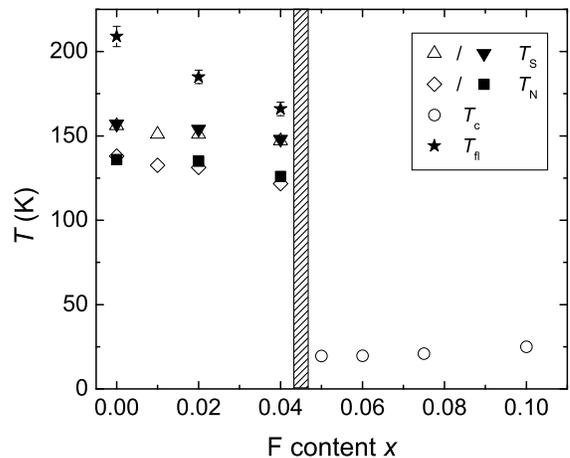}
\end{center}
\caption{Phase diagram of LaFeAsO$_{1-x}$F$_x$ showing the change
of the characteristic temperatures with F content $x$. The onset
of superconductivity in the electrical resistivity at
$T_\mathrm{c}$,\cite{Hess2008} and the temperatures of the
orthorhombic distortion, $T_\mathrm{S}$, and magnetic ordering,
$T_\mathrm{N}$, determined from magnetic susceptibility
\cite{Klingeler2008} and $\mu$SR data \cite{Luetkens2009} are
marked with open symbols. Closed symbols label the characteristic
temperatures determined from thermal expansion measurements in
this work. The striped region highlights the abrupt change from an
orthorhombic/magnetic to a tetragonal/superconducting ground
state.} \label{phase}
\end{figure}

The linear thermal expansion coefficient turned out a very
sensitive probe for the phase transitions in LaFeAsO$_{1-x}$F$_x$
($x\leq 0.04$). Large positive and negative anomalies are found in
$\alpha(T)$ at $T_\mathrm{S}$ and $T_\mathrm{N}$, respectively. By
contrast, the small changes of the unit-cell volume related to
$T_\mathrm{S}$ and $T_\mathrm{N}$ cannot be resolved directly from
XRD measurements. Nevertheless, the close agreement between the
$V_\mathrm{uc}(T)$ curves for LaFeAsO determined from XRD and
$\alpha(T)$ shown in the inset to Fig.~\ref{vol}a confirms the
reliability of our thermal expansion data. The sizeable jumps in
$\alpha$ at $T_\mathrm{N}$ reflect the strong coupling of the
magnetic transition to the crystal lattice. The shape of the
anomalies deviates from the one expected for second-order phase
transitions, which is attributed to the closeness of the
transitions and a contribution from fluctuations above \ts. So
far, the origin of these fluctuations is unknown. A
straightforward interpretation is to attribute them to a competing
instability in vicinity to the actual ground state. One might
speculate that this instability is of magnetic origin as suggested
in Ref.~\onlinecite{Lorenzana2008}. Based on a Hartree-Fock
approximation and a Landau theory, Lorenzana \etal\ find an
orthomagnetic phase competing to the modulated magnetic stripes
which are experimentally observed.~\cite{Lorenzana2008} In this
scenario, long range order of the competing, possibly magnetic
phase is hindered by the orthorhombic distortion, whereas the
increase of the corresponding anomalous positive contribution to
the thermal expansion coefficient is truncated by the structural
transition at \ts. Another model that accounts for the anomalous
thermal expansion coefficient above \ts~comprises ferro-orbital
ordering accompanied by a lattice distortion at \ts.
\cite{Lee2009} In this picture, the enhanced $\alpha$ is suggested
to arise from short-range orbital correlations above \ts.
Interestingly, the experimentally observed onset temperature of
the short range order strongly depends on the F-content. As
visible in Fig.~\ref{phase}(b), the fluctuation regime is much
stronger suppressed for larger $x$ than \tn~and \ts. Further
investigations are necessary to determine the exact nature of the
fluctuations above \ts.

In contrast to the structural and SDW transitions, the
superconducting transitions of LaFeAsO$_{1-x}$F$_x$ with $x =
0.05$ and 0.1 are not seen in $\alpha(T)$. The expected magnitude
of $\Delta \alpha$ can be estimated from Eq.~\ref{eq2} using
literature data. No anomaly was observed at $T_{\rm{C}}$ in the
specific heat of LaFeAsO$_{1-x}$F$_x$.~\cite{Mu2008} As a rough
estimate we take the difference of the curves measured in 0~T and
9~T on LaFeAsO$_{0.9}$F$_{0.1-\delta}$ of $\Delta c_p/T \approx$ 5
mJ/mole K$^2$. The derivative $\partial T_{\mathrm{c}}/{\partial
p}|_{p=0}$ of LaFeAsO$_{0.89}$F$_{0.11}$ was found to be of the
order of 3~K/GPa.~\cite{Takahashi2008} Thus, the anomaly in
$\alpha$ is estimated to $\Delta \alpha \approx 3 \times
10^{-8}$/K. This value, which is too small to be resolved with our
setup, is in line with the absence of large anomalies in $\alpha
(T)$ at $T_{\rm{C}}$. However, it should be noted, that we measure
a directional average of the coefficient of linear thermal
expansion on our polycrystalline samples which is associated to
the hydrostatic pressure dependence. An almost complete
cancellation of the contributions for different crystallographic
directions was observed on single crystals with the related
ThCr$_2$Si$_2$ structure, namely
Ba(Fe$_{0.88}$Co$_{0.12}$)$_2$As$_2$.~\cite{F5-08-2} Measurements
on single crystals of LaFeAsO$_{1-x}$F$_x$ are therefore necessary
to decide, whether the small magnitude of $\Delta \alpha$ is due
to a similar effect.

The superconducting ground state in the $x = 0.05$ sample is
formed at the expense of an abrupt suppression of the structural
and magnetic phase transitions observed in samples with $x \leq
0.04$. The change from the magnetically ordered to the
superconducting ground state upon F doping has been proposed to be
abrupt first-order-like.~\cite{Luetkens2009,Hess2008} Our thermal
expansion data clearly support this picture. While fluctuations
give rise to an enhanced thermal expansion coefficient over an
extended $T$ region for the sample with $x = 0.04$, no indication
of fluctuations is found for $x = 0.05$. Instead, the thermal
expansion coefficient of this sample is almost identical to the
one for $x = 0.1$, which lies well in the superconducting regime
of the phase diagram. Moreover, the disappearance of the
fluctuation regime around $x = 0.045$ cannot be explained by a
smooth convergence of $T_\mathrm{fl}$ and \ts~with increasing F
content, as seen from the phase diagram Fig.~\ref{phase}b.
Therefore, our data corroborate a first-order-like scenario for
the transition towards superconductivity upon doping.


\section{Summary}

We performed measurements of the linear thermal expansion
coefficient of LaFeAsO$_{1-x}$F$_x$ with $x \leq 0.1$ in the
temperature range between 5~K and 300~K. The structural and SDW
transitions of the compounds with low F content $x \leq 0.04$ give
rise to large anomalies in $\alpha(T)$, whereas fluctuations are
present also well above $T_\mathrm{S}$. By contrast, the
superconducting transitions of the samples with $x \geq 0.05$ are
not observable. Moreover, no indications for residual fluctuations
in the superconducting samples are found at elevated temperatures,
not even at the lowest F content $x = 0.05$. This finding supports
the idea of an abrupt first-order type transition between the
magnetic and superconducting ground states upon fluorine
substitution.

\begin{acknowledgments}
We thank M. Deutschmann, S. M\"uller-Litvanyi, R. M\"uller, J.
Werner, and S. Ga{\ss} for technical support. This work has been
supported by the Deutsche Forschungsgemeinschaft, through
BE1749/12 and through FOR 538 (BU887/4).
\end{acknowledgments}

\bibliographystyle{unsrtnat}
\bibliography{LaFeAsO_TE_UK}

\end{document}